# A Framework for IoT-Enabled Smart Manufacturing for Energy and Resource Optimization

## Bazigu Alex* [1,2] and Mwebaze Johnson[1]

[1]College of Computing and Information Sciences, Makerere University, Kampala, Uganda

[2]Faculty of Science and Technology, Victoria University Kampala, Uganda

*Corresponding Author: bazigu.alex@gmail.com, Tel.: +256772098106

## ABSTRACT

The increasing demands for sustainable and efficient manufacturing systems have driven the integration of Internet of Things (IoT) technologies into smart manufacturing. This study investigates IoT-enabled systems designed to enhance energy efficiency and resource optimization in the manufacturing sector, focusing on a multi-layered architecture integrating sensors, edge computing, and cloud platforms.

MATLAB Simulink was utilized for modeling and simulation, replicating typical manufacturing conditions to evaluate energy consumption, machine uptime, and resource usage. The results demonstrate an 18% reduction in energy consumption, a 22% decrease in machine downtime, and a 15% improvement in resource utilization. Comparative analyses highlight the superiority of the proposed framework in addressing operational inefficiencies and aligning with sustainability goals.

The study underscores the potential of IoT in transforming traditional manufacturing into interconnected, intelligent systems, offering practical implications for industrial stakeholders aiming to optimize operations while adhering to global sustainability standards. Future work will focus on addressing identified challenges such as high deployment costs and data security concerns, aiming to facilitate the broader adoption of IoT in industrial applications.

# 1. INTRODUCTION

The advent of Industry 4.0 has revolutionized manufacturing by integrating cutting-edge technologies such as the Internet of Things (IoT) (Suresh et al., 2014), artificial intelligence, and big data analytics (Jagatheesaperumal et al., 2022; Vikranth & Prasad, 2021; Zheng et al., 2024). Despite challenges, these innovations enable real-time data exchange, predictive analytics, and autonomous decision-making (Zheng et al., 2024), facilitating the shift from traditional to smart manufacturing systems (Antony et al., 2020; Aouedi & Piamrat, 2024). In this rapidly evolving environment, energy efficiency and resource optimization have become critical for enhancing operational sustainability, particularly in regions with limited infrastructure and high production costs, such as Uganda (Antony et al., 2020).

IoT, initially conceptualized by Kevin Ashton in 1999, connects machines, billions of devices such as sensors, actuators and systems within a unified network, fostering seamless communication and data-driven insights (Yang et al., 2019). This capability has positioned IoT as a cornerstone of Industry 4.0, enabling manufacturers to monitor energy consumption, identify inefficiencies, and automate corrective measures (Xu et al., 2018). For instance, IoT-based energy management systems can reduce energy costs by up to 20%, as demonstrated in energy-intensive industries (Reichardt et al., 2024).

The effectiveness of IoT in manufacturing lies in its three-layer architecture: sensing, data transmission, and analytics. Sensors gather data on critical parameters such as energy usage and machine performance, transmitting it to centralized platforms or cloud systems for analysis as showed in Fig. 1. Advanced analytics and machine learning algorithms generate actionable insights that improve decision-making and operational efficiency (Jayakumar et al., 2021; Soori et al., 2023).

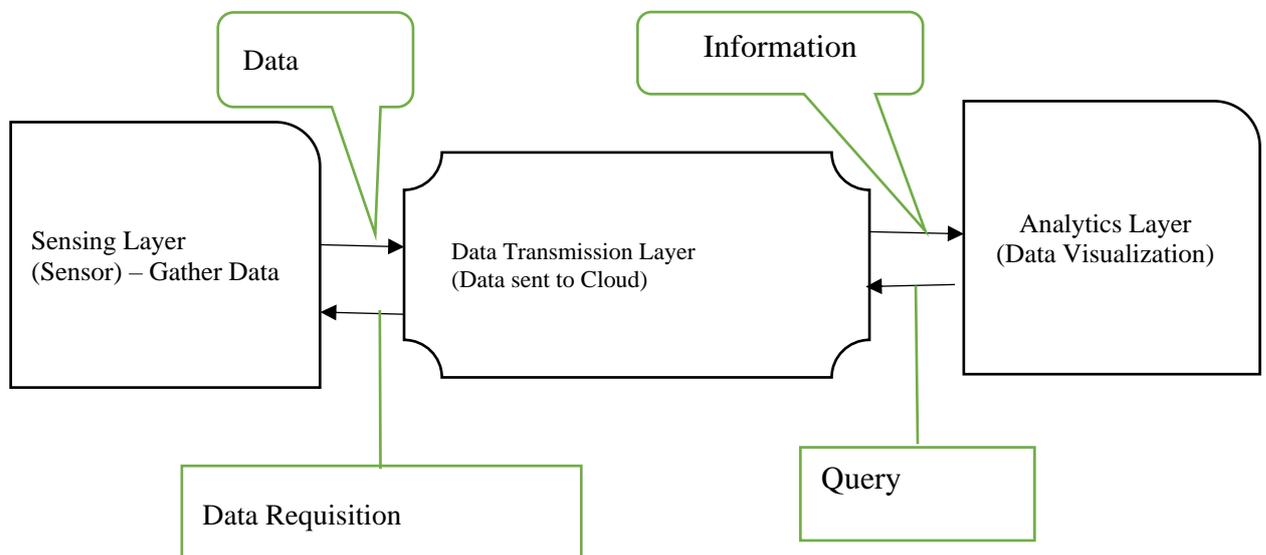

Fig. 1. Components of IoT

In Uganda, the industrial sector, which includes manufacturing, mining, construction, and utilities, contributed approximately 26.77% in 2022 to the GDP (Statista, 2022), of this, manufacturing alone accounted for approximately 15.65% as of 2023 (World Bank, 2023), However, energy costs continue to impose a substantial operational challenge for the sector (Africa Press, 2024; Private Sector Foundation Uganda, 2024). Kakira Sugar Limited, one of Uganda's largest sugar producers, faces challenges in optimizing energy usage and resource management, despite using its co-generation facility to produce electricity from bagasse that partially addresses its energy needs (Adeyemi & Asere, 2014). The continued reliance on biomass and limited diversification of energy sources contribute to inefficiencies and highlight the need for improved energy management practices (Antony et al., 2020). Despite infrastructural limitations typical of developing economies (Gurara et al., 2018), IoT-driven systems provide opportunities for cost-effective energy monitoring and optimization, bridging the gap between traditional practices and modern energy-saving technologies (Antony et al., 2020). With IoT adoption, industries like Kakira Sugar Limited can significantly mitigate operational inefficiencies while aligning with global energy sustainability goals enhance competitiveness (Gasper et al., 2019; Hales & Birdthistle, 2022).

Driven by the need to address inefficiencies in energy use and resource management, this study explored the potential of IoT to optimize energy efficiency and resource utilization in Uganda's manufacturing sector. Existing IoT frameworks often depend on advanced infrastructure and high costs, making them unsuitable for resource-constrained environments like Uganda. Key gaps include reliance on expensive infrastructure, difficulty integrating with legacy systems, and inadequate cybersecurity measures (Cheikh et al., 2022; Kaur & Sood, 2020; Wang et al., 2016). These shortcomings highlight the need for solutions tailored to environments with financial and infrastructural constraints. Uganda's manufacturing sector presents unique challenges compared to other regions, including heavy dependence on biomass energy, limited energy diversification, and significant infrastructural barriers (Adeyemi & Asere, 2014; Antony et al., 2020). For instance, unlike industries in more developed regions with access to diversified energy sources, manufacturers in Uganda, such as Kakira Sugar Limited, rely on co-generation facilities powered by byproducts like bagasse. Despite these efforts, inefficiencies persist due to the lack of real-time energy monitoring and predictive maintenance capabilities (Kagira et al., 2020).

To address these challenges, this study introduces a tailored IoT framework that emphasizes affordability, adaptability to existing systems, and real-time monitoring of energy and resources. The framework is generalizable to other resource-constrained manufacturing environments, with Kakira Sugar Limited used as a case study to demonstrate its applicability. The analysis is guided by the Technology Acceptance Model (TAM) (Silva, 2015), Diffusion of Innovations theory (Rogers et al., 2014), and the Triple Bottom Line (TBL) framework to align with economic, environmental, and social sustainability goals (Isil & Hernke, 2017).

The rest of this paper is structured as follows: Section 2 reviews related work on IoT-enabled smart manufacturing and energy optimization. Section 3 details the methodology, including system requirements, IoT model development, and simulation approach. The proposed system architecture, supported by diagrams and technical specifications presented in Section 4, followed by results in Section 5 and discussion while aligning with the proposed framework and the research is concluded with the challenges, key findings, contributions and future research directions in Section 6 and 7.

## 2. RELATED WORK

IoT technologies have profoundly transformed various sectors, including healthcare, transportation, and manufacturing. In healthcare, IoT facilitates remote patient monitoring, reducing hospital readmission rates by up to 60%. In transportation, connected systems improve fuel efficiency by approximately 20%. In manufacturing, IoT enhances operational performance through predictive maintenance, which can reduce downtime by up to 50%, while also optimizing energy consumption and resource utilization (Kunwar, 2024).

The evolution of industrialization, culminating in Industry 4.0, has introduced advanced technologies such as IoT, artificial intelligence, and big data analytics. These innovations enable real-time monitoring, predictive analytics, and autonomous decision-making in manufacturing systems, significantly improving efficiency and sustainability (Ashton, 1997; Mokyr & Strotz, 1998; Rifkin, 2012; Schwab, 2016; Xu et al., 2018).

In smart manufacturing, IoT addresses critical challenges, including high energy costs and resource inefficiencies. Studies emphasize its ability to facilitate real-time data acquisition, machine monitoring, and process optimization, thereby minimizing energy wastage and improving resource utilization (Jamwal et al., 2021; Xu et al., 2018). Smart grids and IoT-based energy management systems improve cost efficiency and reduce energy consumption (Xu et al., 2018). Studies highlight IoT's role in optimizing production, minimizing energy wastage, and enhancing resource efficiency by tracking raw material usage and reducing delays (Jamwal et al., 2021; Yang et al., 2019). However, high investment costs, cybersecurity risks, and integration challenges with legacy systems remain significant barriers, particularly in resource-constrained environments (Soori et al., 2023). Developing cost-effective and secure frameworks is essential for broader IoT adoption in such contexts.

Energy efficiency has been a key focus of IoT applications in manufacturing. Solutions such as "Self-Organized Things" (SoT) and energy-efficient indexing systems like EGF-tree have been developed to conserve sensor energy through dynamic sleep modes and hierarchical data organization (Aouedi & Piamrat, 2024; Kagira et al., 2020, 2020; Khan et al., 2023; Lombardi et al., 2021; Pethe et al., 2024). Tailored frameworks, such as those applied in Uganda's sugar industry, demonstrate the potential for IoT to enhance energy tracking, predictive maintenance, and overall energy efficiency, addressing the unique challenges faced by industries in resource-constrained settings (Kagira et al., 2020; A. Kaur & Sood, 2020).

IoT has also been instrumental in advancing resource optimization by enabling precise real-time monitoring of raw materials, water consumption, and labor. In industries such as sugar production, IoT-based systems have been used to optimize byproduct utilization, such as bagasse and molasses, aligning with circular economy principles and sustainability goals (Ahmad et al., 2024; Gubbi et al., 2013; Reichardt et al., 2024). However, widespread implementation is hindered by challenges related to affordability, limited technical expertise, and interoperability between devices, particularly in regions with infrastructural and financial constraints (Antony et al., 2020; Lombardi et al., 2021).

Several IoT frameworks have been proposed to address energy efficiency and resource optimization in manufacturing. Green industrial IoT architectures and edge-computing models have shown promise in enhancing operational efficiency but are often reliant on advanced infrastructure, limiting their applicability in developing regions (Isil & Hernke, 2017; Singh & Sharma, 2021). Multi-layered frameworks, such as LoRaWAN, provide scalable and energy-efficient solutions but face challenges related to data security and integration with renewable energy systems (Silva, 2015; Singh & Sharma, 2021). These gaps highlight the need for adaptive, cost-effective IoT solutions tailored to the specific operational constraints of industries in resource-constrained environments.

## 3. SYSTEM DESIGN

This study proposes an IoT-enabled smart manufacturing system to enhance energy efficiency and optimize resource utilization within the manufacturing sector. The hardware components consisted of sensors, actuators, and gateways. Sensors, such as temperature, pressure, and energy meters, were utilized to monitor machine performance and energy usage. Actuators facilitated real-time machinery control based on sensor feedback, while gateways aggregated sensor data and transmitted it to cloud or edge servers for further processing. Lightweight and efficient communication protocols, including Message Queue Telemetry Transport (MQTT) and Constrained Application Protocol (CoAP), were employed to ensure seamless data transmission. At the same time, edge computing frameworks enable localized data analysis, minimizing latency and supporting real-time decision-making.

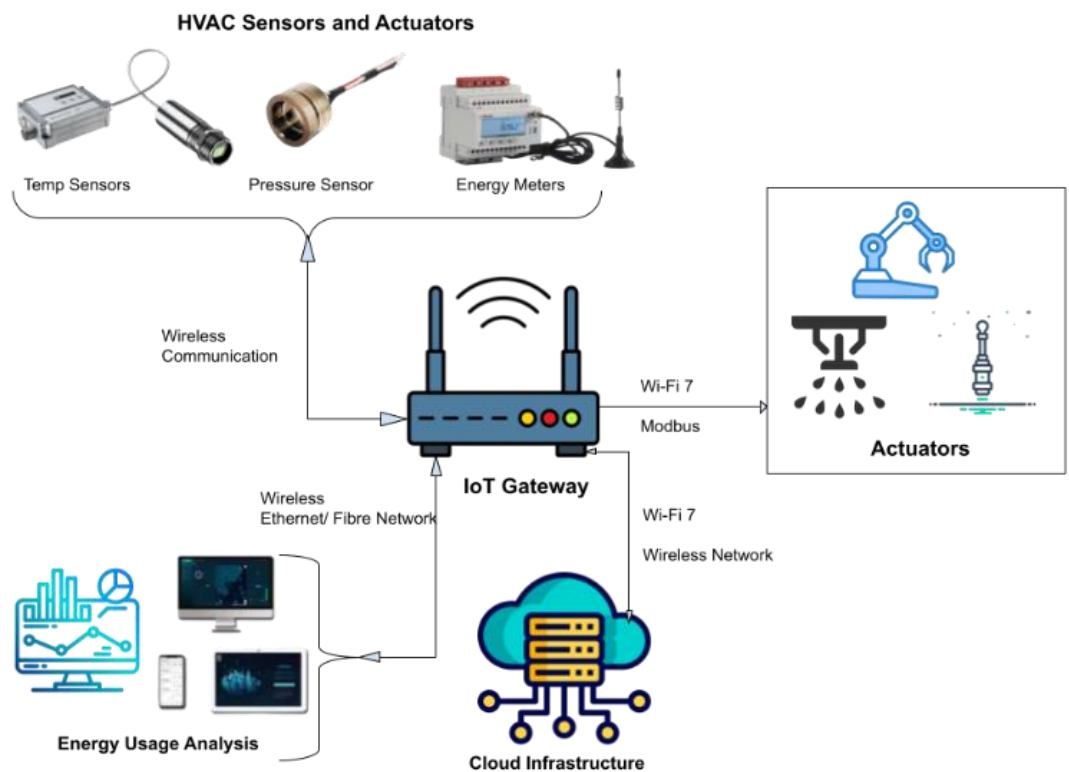

Fig.2. Block Diagram for the IoT Enabled Smart Manufacturing Architecture [1].

The proposed IoT model integrates these hardware and software components to simulate real-world manufacturing scenarios. Sensors collect data on energy consumption, machine status, and environmental conditions. This data is transmitted through gateways to a cloud platform, where advanced analytics and machine learning algorithms process it. The model detects anomalies, optimizes system operations, and minimizes downtime through predictive maintenance.

The IoT model is validated using MATLAB Simulink, a powerful tool for dynamic system modeling and analysis (D'Angelo et al., 2017). The simulation focuses on evaluating energy consumption patterns, machine efficiency, and resource utilization. The results provide empirical evidence of the framework's effectiveness in meeting the study's objectives.

## 3.1 System Architecture

The system architecture follows a multi-layered approach to ensure efficient data flow and operational effectiveness. It comprises the perception layer, responsible for data acquisition through sensors and actuators; the network layer, which facilitates data transmission using communication protocols such as Wi-Fi, ZigBee, and LoRaWAN; the processing layer, consisting of edge devices and cloud platforms for data preprocessing, storage, and analysis; and the application layer, which provides user interfaces for monitoring and controlling manufacturing processes (Lombardi et al., 2021), accessible on both mobile and desktop platforms.

## 4. SIMULATION

The IoT - enabled simulation model (Fig. 3) was developed using MATLAB Simulink to test and validate the proposed framework.

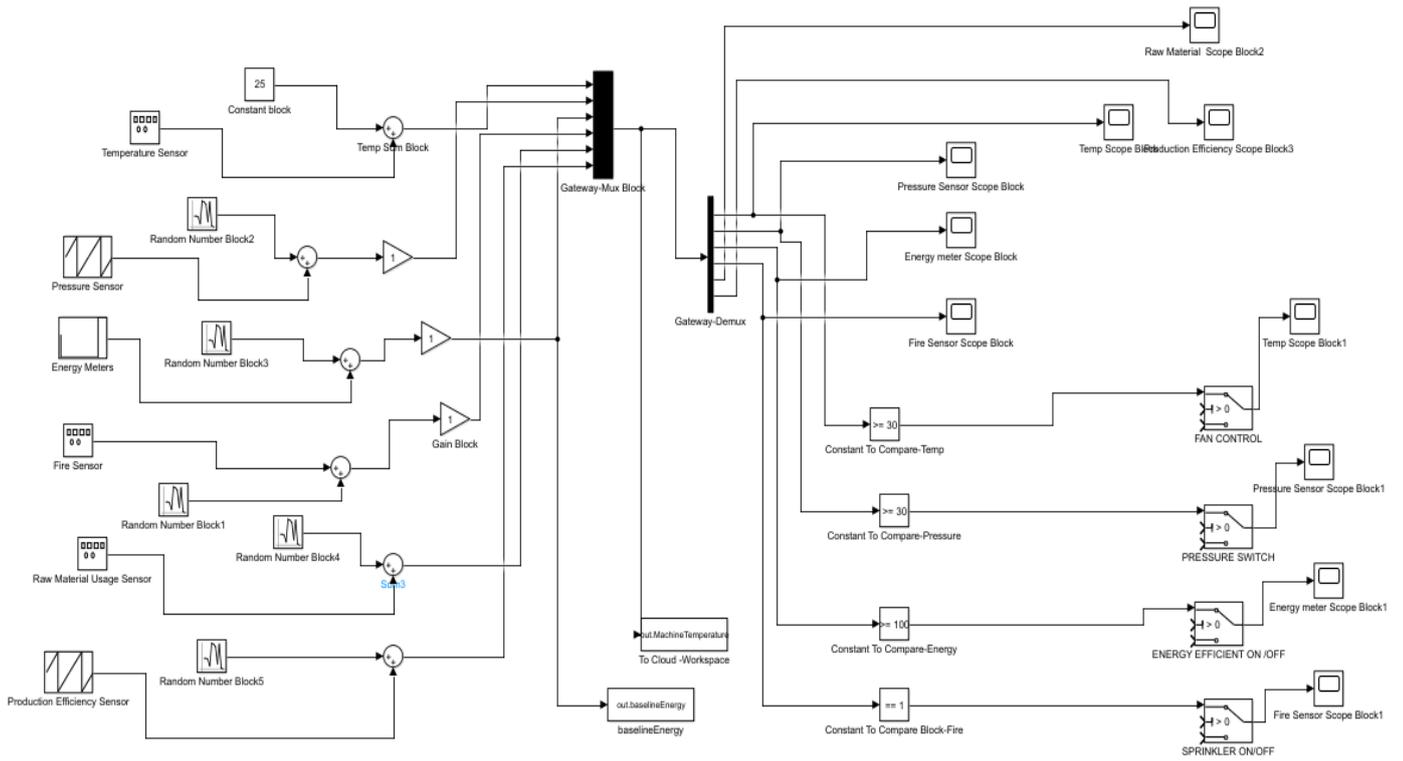

Fig. 3. IoT Enabled Simulation Model

The model integrates sensors to monitor critical parameters such as energy usage, temperature, pressure, and machine performance. Data collected by the sensors was preprocessed by edge devices to reduce transmission latency and improve data quality. The data was transmitted to cloud servers for in-depth analysis. Predictive maintenance algorithms were implemented to optimize operations and minimize resource wastage.

The technical setup included sensor calibration for accurate measurement, such as energy usage (±0.1%) and temperature (±0.5°C). A star topology was adopted for efficient data transmission, with MQTT used for real-time communication and HTTP for periodic data transfers. The simulation replicated a typical manufacturing environment, including machines equipped with IoT sensors, gateways, and edge devices. Key input variables, such as energy consumption rates, machine runtime, and ambient temperature, were systematically varied to evaluate system performance.

Data for both baseline and optimized scenarios were simulated to assess energy consumption, machine uptime, and material usage. Baseline data represented traditional manufacturing setups, while optimized data reflected the effects of predictive maintenance and real-time energy management. Simulation outputs were analyzed in MATLAB to calculate efficiency improvements and resource optimization metrics. The simulation parameters were based on typical industrial conditions, as outlined in (N. Kaur & Sood, 2017). The results align with empirical findings in studies such as (Aouedi & Piamrat, 2024)

## 5. RESULTS

The simulation revealed significant improvements in key performance metrics, as illustrated in the following sections:

### 5.1 Energy Reduction

he IoT-enabled system reduced energy consumption by 18% compared to baseline scenarios. This improvement resulted from optimized machine operations and real-time energy adjustments, as shown in Fig. 4.

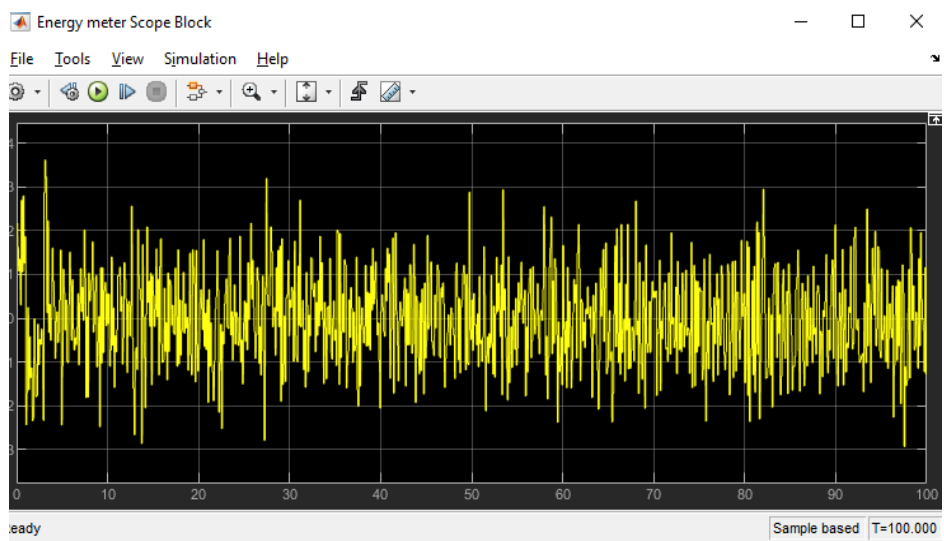

Fig. 4. Energy Meter Scope Block

The system effectively minimized energy waste by identifying idle periods and shutting down non-essential components.

### 5.2 Improved Machine Uptime

Continuous monitoring and predictive maintenance reduced machine downtime by 22%. Early anomaly detection prevented sudden equipment failures, ensuring uninterrupted production. Fig. 5 demonstrates stable temperature regulation, which enhanced machine reliability and minimized thermal wear.

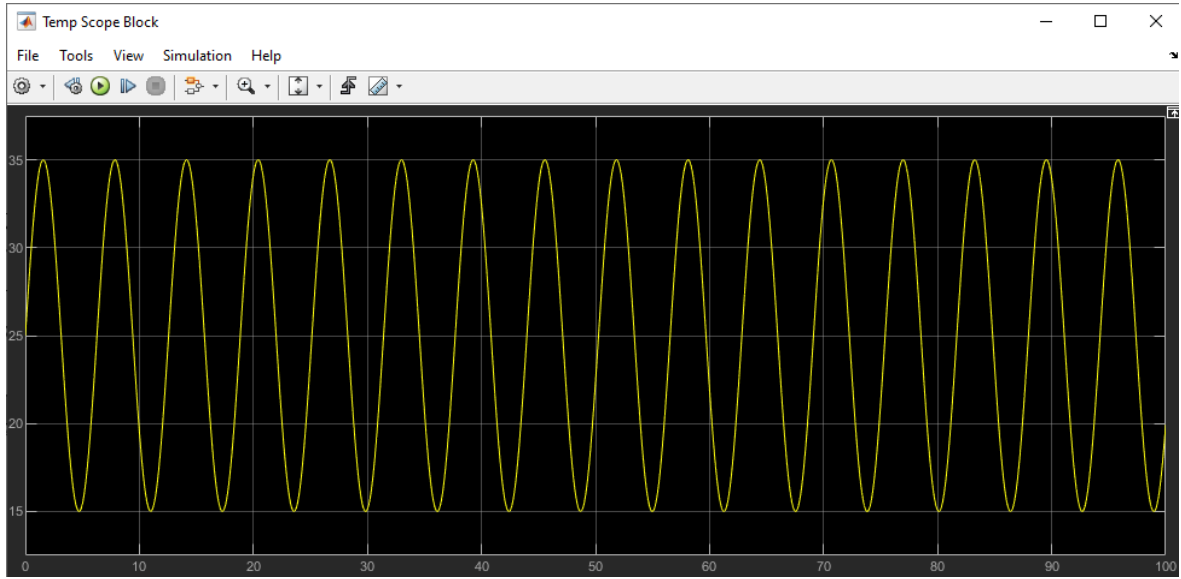

Fig. 5. Temperature Sensor Scope Block

This graph shows stable temperature trends with brief spikes during high workloads. The system's cooling mechanisms activated promptly, preventing overheating and improving component lifespan.

### 5.3 Resource Optimization

Material waste decreased by 15%, as shown in Fig. 6, due to real-time monitoring and adjustments in production schedules.

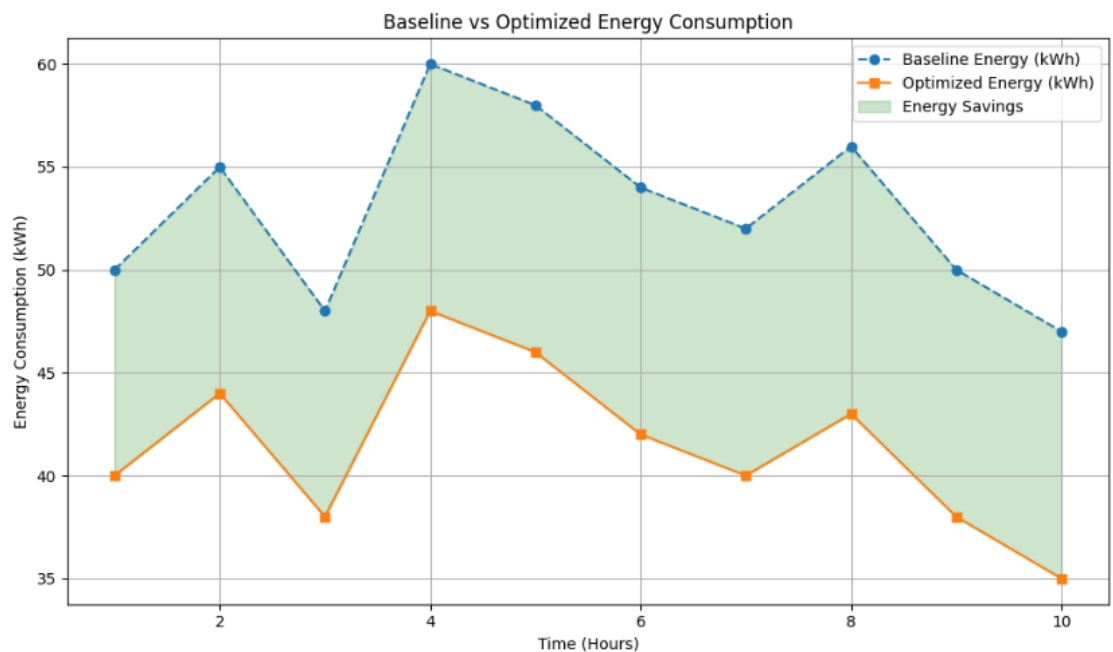

Fig. 6. Baseline Vs Optimized Energy Consumption.

Fig. 6 demonstrates consistent energy savings across all machines, validating the system's scalability and effectiveness in diverse manufacturing environments.

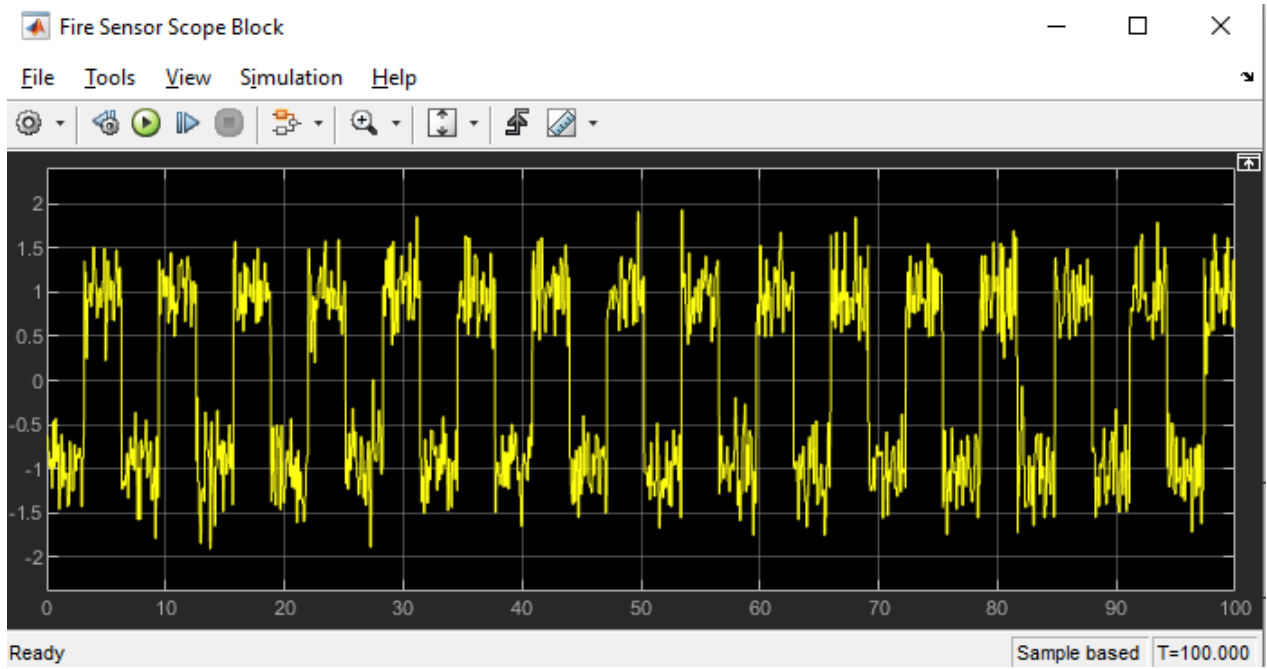

Fig. 6. Fire Sensor Scope Block

Fire sensor readings remained stable under normal conditions, with occasional spikes during test scenarios. The rapid response of the sprinkler system ensured safety and reduced downtime risks.

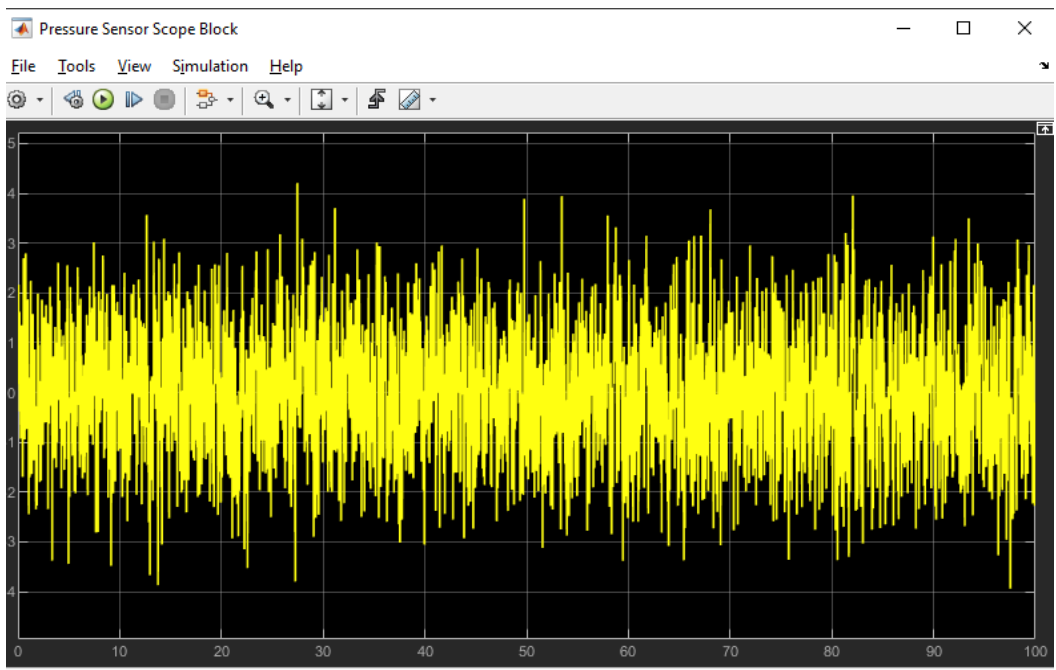

Fig. 7. Pressure Sensor Scope Block

Pressure variations were well-contained, demonstrating effective regulation during production cycle transitions. Optimal pressure levels reduced equipment wears and energy waste.

## 6. DISCUSSION

The simulation results demonstrate the effectiveness of the IoT-enabled smart manufacturing system in improving energy efficiency, machine uptime, and resource utilization. Real-time data acquisition and predictive analytics enabled precise monitoring and control, leading to streamlined operations. The system's capacity to detect anomalies and optimize production aligns with prior research and demonstrates its relevance in industrial applications. The comparative analysis between baseline and optimized scenarios highlights significant operational improvements and validates the practical benefits of IoT integration.

However, several challenges were encountered during implementation. High initial costs for advanced sensors and edge devices could hinder adoption in smaller enterprises. Integrating legacy systems required custom communication protocols, increasing complexity. Additionally, data security concerns arose due to the transmission of sensitive information to cloud servers. Scaling the system for larger manufacturing setups also led to occasional latency issues, suggesting the need for further optimization of network and processing infrastructures. Addressing these challenges will be critical to achieving widespread adoption of IoT-enabled smart manufacturing systems.

## 7. CONCLUSION

This study successfully demonstrated the potential of IoT-enabled smart manufacturing systems in improving energy efficiency, machine uptime, and resource utilization. By integrating advanced sensors, real-time data analytics, and predictive maintenance, the system achieved an 18% reduction in energy consumption, a 22% decrease in machine downtime, and a 15% reduction in material waste. These findings highlight the system's capacity to enhance operational efficiency and sustainability in manufacturing. Future research could explore cost-effective sensor alternatives, advanced security protocols, and scalability optimizations to further strengthen the adoption of IoT in industrial environments.